\documentclass[prd,twoside,preprintnumbers,superscriptaddress,nofootinbib]{revtex4-1}

\usepackage{amsmath,amssymb,slashed}
\usepackage{dsfont}
\usepackage{graphicx}
\usepackage{epstopdf}  
\usepackage{hyperref}

\begin{document}

\title{Goldberger-Treiman relation and Wu-type experiment in the decuplet sector}

\author{Magnus Bertilsson}
\author{Stefan Leupold}

\affiliation{Institutionen f\"or fysik och astronomi, Uppsala universitet, Box 516, S-75120 Uppsala, Sweden}
\date{\today}

\begin{abstract}
  The leading-order chiral Lagrangian for the baryon octet and decuplet states coupled to Goldstone bosons and external sources 
  contains six low-energy constants. Five of them are fairly well known from phenomenology, but the sixth one is 
  practically unknown. This coupling constant
  provides the strength of the (p-wave) coupling of Goldstone bosons to decuplet states. Its size and even sign are under debate. 
  Quark model and QCD for a large number of colors provide predictions, but some recent phenomenological analyses suggest even 
  an opposite sign for the Delta-pion coupling. The Goldberger-Treiman relation connects this coupling constant to the axial 
  charge of the Delta baryon. This suggests a Wu-type experiment to determine the unknown low-energy constant. While this 
  is not feasible in the Delta sector because of the large hadronic width of the Delta, there is a flavor symmetry 
  related process that is accessible: the weak semileptonic
  decay of the Omega baryon to a spin 3/2 cascade baryon. A broad research program is suggested that can pin down at 
  least the rough size and the sign 
  of the last unknown low-energy constant of the leading-order Lagrangian. It encompasses experimental measurements, 
  in particular the forward-backward asymmetry of the semileptonic decay, together with a determination of the quark-mass 
  dependences using lattice QCD for the narrow decuplet states and chiral perturbation theory to extrapolate to the Delta sector.
  Besides discussing the strategy of the research program, the present work provides a feasibility check based 
  on a simple leading-order calculation. 
  \end{abstract}

\maketitle

\section{Motivation}

The strong interaction binds quarks to hadrons and protons and neutrons to atomic nuclei. It is generally accepted that 
Quantum Chromodynamics (QCD) provides the basis for a quantitative description of all phenomena governed by the strong 
interaction \cite{pesschr}. Yet, understanding the non-perturbative aspects of QCD provides the most challenging problem 
within the standard model of particle physics. 

Since the famous prediction of the $\Omega$ baryon based on the SU(3) flavor symmetry proposed by 
Gell-Mann \cite{Gell-Mann:1961omu,Gell-Mann:1962yej} and Ne'eman \cite{Neeman:1961jhl}, 
the flavor-decuplet states have been very instrumental in 
revealing properties of the strong interaction and therefore of QCD. Even if one restricts the attention to the 
ground-state baryon octet, with the proton and neutron (nucleons) as its most prominent members, 
the dynamics of QCD involves the $\Delta$ and other 
decuplet states at the quantum level, since those are close in mass to the nucleons and couple strongly. 
Examples for the importance of decuplet states range from weak decays of octet 
baryons \cite{Jenkins:1991es,Jenkins:1991bt,Flores-Mendieta:2019lao,Alvarado:2021ibw} 
over baryon radii \cite{Granados:2016jjl,Alarcon:2017asr,Granados:2017cib,Leupold:2017ngs,Alvarado:2023loi} 
and pion-nucleon scattering \cite{Yao:2016vbz} 
to the properties of atomic nuclei 
and nuclear reactions \cite{Ericson:1988gk,Alvarez-Ruso:1998ais,Krebs:2007rh,Buss:2011mx,NuSTEC:2017hzk}. 

Chiral perturbation theory constitutes the effective field theory of QCD in the low-energy sector of light quarks 
\cite{Weinberg:1978kz,Gasser:1984gg,Jenkins:1991es,Becher:2001hv,Scherer:2012xha}. 
In the baryon sector, 
the two lowest-lying multiplets (octet and decuplet) are separated by a mass gap from the higher-lying states. In the quark 
model, this gap is caused by the excitation of orbital angular momentum. In contrast, the change from the octet to the decuplet
requires only a spin flip; see, e.g., the minireview of the quark model in \cite{pdg}. 
In QCD for a large number $N_c$ of quark colors \cite{tHooft:1973alw,Witten:1979kh}, the mass difference between 
octet and decuplet states vanishes \cite{Dashen:1993jt} (in the chiral limit). 
The bottom line of these considerations is that there 
is an energy regime where it makes sense 
to consider an effective field theory with the Goldstone bosons, the octet baryons and the decuplet baryons as relevant 
degrees of freedom. 
The leading-order chiral Lagrangian 
including the decuplet states 
is given by \cite{Jenkins:1991es,Lutz:2001yb,Semke:2005sn,Pascalutsa:2006up,Ledwig:2014rfa,Holmberg:2018dtv,Mommers:2022dgw} 
\begin{eqnarray}
  && {\cal L}_{\rm baryon}^{(1)} = {\rm tr}\left(\bar B \, (i \slashed{D} - m_{(8)}) \, B \right)  \nonumber \\ 
  && {}+ \bar T_{abc}^\mu \, ( i \gamma_{\mu\nu\alpha} (D^\alpha T^\nu)^{abc} - \gamma_{\mu\nu} \, m_{(10)} \, (T^\nu)^{abc})
  \nonumber \\ 
  && {}+ \frac{D}{2} \, {\rm tr}(\bar B \, \gamma^\mu \, \gamma_5 \, \{u_\mu,B\}) 
  + \frac{F}{2} \, {\rm tr}(\bar B \, \gamma^\mu \, \gamma_5 \, [u_\mu,B])  \nonumber \\
  && {} + \frac{h_A}{2\sqrt{2}} \, 
  \left(\epsilon^{ade} \, \bar T^\mu_{abc} \, (u_\mu)^b_d \, B^c_e
    + \epsilon_{ade} \, \bar B^e_c \, (u^\mu)^d_b \, T_\mu^{abc} \right) \nonumber \\
  && {} -\frac{H_A}{2} \, \bar T^\mu_{abc} \gamma_\nu \gamma_5 \, (u^\nu)^c_d \; T_\mu^{abd}   \,.
  \label{eq:baryonlagr}
\end{eqnarray}
For a detailed definition of all the building blocks see \cite{Holmberg:2018dtv,Mommers:2022dgw}. What matters here 
are the following six low-energy constants.

\underline{$m_{(8)}$ and $m_{(10)}$:} These are the masses for the baryon octet and decuplet, respectively, in the chiral limit.
Their values are essentially known from the Gell-Mann--Okubo mass relations \cite{Gell-Mann:1961omu,Okubo:1961jc}. 
Yet, if one aims for higher accuracy, e.g.\ for the nucleon, the determination of the 
key quantity, the nucleon sigma term, is a matter of active research \cite{RuizdeElvira:2017stg}. 

\underline{$D$ and $F$:} These are the axial charges of the octet baryons. They are essentially known from 
semileptonic weak decays of the octet 
baryons \cite{Cabibbo:1963yz,Jenkins:1991es}. Also here, various methods for a more precise determination 
(chiral perturbation theory, lattice QCD) are currently developed 
\cite{Ledwig:2014rfa,RQCD:2019jai,Sauerwein:2021jxb,Bali:2023sdi}. 

\underline{$h_A$:} This interaction strength connecting baryon octet and decuplet is essentially known from the 
  strong decays of decuplet baryons to octet baryons and pions. If one fits a 
  leading-order calculation based on \eqref{eq:baryonlagr} to the data, the values for $h_A$ obtained from the different 
  strangeness sectors show a spread of about $\pm20$\%  (see, e.g., \cite{Dashen:1994qi,Holmberg:2019ltw}). 
  This is the expected size of flavor-symmetry breaking. 

\underline{$H_A$:} This low-energy constant provides the strength of the coupling of decuplet states to Goldstone bosons.
  On account of the Goldberger-Treiman relation \cite{Goldberger:1958tr,Goldberger:1958vp} 
  it might be called axial charge of the decuplet states (in the chiral limit). 
  This quantity is to a large extent unknown. 

The purpose of the present work is to devise a strategy that can lift our knowledge of $H_A$ to the level of the other 
leading-order low-energy constants.

\section{Key observable}

Theory estimates for $H_A$ exist. What is missing, however, are a direct experimental determination and a 
first-principle calculation based, e.g., on lattice QCD. If one considers two-flavor QCD for a large number of quark colors
(large-$N_c$ limit), 
one obtains $H_A = \frac95 \, g_A  \approx 2.27$ \cite{Pascalutsa:2006up,Ledwig:2011cx} with the axial charge $g_A = D+F$ of the 
nucleon. This agrees also with a quark-model 
estimate \cite{Unal:2021byi}.\footnote{In \cite{Unal:2021byi,Yao:2016vbz}, the low-energy constant $H_A$ is called $g_1$.} 
Three-flavor QCD for a large number of quark colors 
yields $H_A = 9F -3D \approx 1.74$ \cite{Dashen:1994qi,Semke:2005sn}.\footnote{In \cite{Dashen:1994qi}, the low-energy constant $H_A$ is called $-{\cal H}$.} Below, we will use $\vert H_A \vert = 2$ for a sample calculation. 

Note that all these theory estimates claim that $F$, $D$ and $H_A$ have the {\it same sign}. In this context it is important to 
stress that the sign 
of $h_A$ relative to the other low-energy constants is a matter of conventions. But the relative signs between $F$, $D$ and $H_A$
have physical significance. Linear combinations of $F$ and $D$ enter the various formulae for semileptonic hyperon decays and 
baryon-meson coupling constants \cite{Dashen:1993jt}. Linear combinations of $g_A^3 = (F+D)^3$ and $H_A \cdot h_A^2$ 
enter, e.g., the loop corrections of pion-nucleon scattering \cite{Yao:2016vbz}. 

In contrast to the previously discussed ``equal-sign'' estimates, 
a sign of $H_A$ opposite to the quark-model prediction has been suggested by some 
phenomenological analyses based on two-flavor loop calculations in chiral perturbation theory that include the Delta degrees 
of freedom \cite{Yao:2016vbz,Unal:2021byi}. In \cite{Yao:2016vbz} pion-nucleon scattering has been analyzed. Here $H_A$ appears 
in the loop corrections in its incarnation as the strong coupling constant of 
$\Delta$-$\Delta$-$\pi$.\footnote{Strictly speaking, the Goldstone bosons couple in 
two partial waves to the spin-3/2 states: a p- and an f-wave. But the latter can only appear in the 
next-to-next-to-leading-order Lagrangian at the earliest.} 
Naturally the determination of $H_A$ is here very indirect. 

Even for the nucleon-nucleon-pion coupling, a determination via 
pion-nucleon or nucleon-nucleon scattering is more indirect and less precise than the direct determination via the 
axial charge $g_A$. In fact, 
the Goldberger-Treiman relation \cite{Goldberger:1958tr,Goldberger:1958vp}, 
which is inherent to the chiral Lagrangian (\ref{eq:baryonlagr}), relates a strong coupling 
to a weak process. For the nucleon, it relates the $N$-$N$-$\pi$ coupling to the beta decay. For the Delta states this would 
translate to processes like 
\begin{eqnarray}
  \label{eq:Delta-weak}
  \Delta^0 \to \Delta^+ \, e^- \bar\nu_e
\end{eqnarray}
suggesting a Wu-type experiment \cite{Wu:1957my} for Delta baryons. 
However, the pions are lighter than the mass difference between Delta and nucleon. 
Therefore the Delta states are very short-lived. 
Their weak decays like (\ref{eq:Delta-weak}) are not accessible in practice.
However, SU(3) flavor symmetry, which is also inherent to the 
leading-order chiral Lagrangian (\ref{eq:baryonlagr}), relates such inaccessible semileptonic Delta decays to measurable 
$\Omega$ decays. The latter is stable with respect to the strong interaction because the kaon is heavier than the mass difference
between $\Omega$ and the cascade state $\Xi$ \cite{pdg}. 

The flavor sibling of (\ref{eq:Delta-weak}) that is accessible by experiments is 
\begin{eqnarray}
  \label{eq:Omega-weak}
  \Omega^- \to \Xi^{*0} \, \ell^- \bar\nu_\ell \quad \mbox{(Wu-type experiment)}
\end{eqnarray}
where $\Xi^{*0}$ denotes the neutral excited cascade state of spin 3/2 and $\ell$ denotes an electron or muon. The state 
$\Xi^{*0}$ can be completely reconstructed from charged states, certainly an advantage for experimental analyses. The decay 
sequence is $\Xi^{*0} \to \pi^+ \Xi^-$, followed by $\Xi^- \to \Lambda \pi^-$ at a displaced vertex, 
followed by $\Lambda \to \pi^- p$ at another displaced vertex. The unobserved neutrino can be reconstructed by the missing mass. 
It should be stressed, however, that so far 
this process \eqref{eq:Omega-weak} has not been observed. Below we will estimate the branching ratio and 
demonstrate that the observation of the process should be feasible by present-day experiments. 

One might think that the $\Delta$-$\Delta$-$\pi$ coupling constant can be determined from lattice QCD. However, lattice QCD 
faces similar problems as the experiments. For physical quark masses, the Delta states are so broad that they need to be 
reconstructed from pion-nucleon scattering by the L\"uscher method \cite{Luscher:1990ux}. Thus one would have to carry out a 
detailed coupled-channel analysis involving five-point functions with two three-quark currents (baryons) and 
three quark-antiquark currents (mesons). 

Also in lattice QCD, weak processes like (\ref{eq:Omega-weak}), related by the Goldberger-Treiman relation to the coupling 
$\Omega$-$\Xi^*$-$K$, are much more promising to pin down $H_A$. Note that the $\Xi^*$, albeit unstable, has only a width of
$9\,$MeV. A traditional analysis, treating the $\Xi^*$ as if stable \cite{BMW:2008jgk}, should yield reasonable results. Still we 
stress that such results require interpretation as demonstrated for the case of the Goldberger-Treiman relation applied to the 
nucleon \cite{Bali:2018qus}. 

Of course, our world is not SU(3) flavor symmetric. Yet, the traditional success of Gell-Mann and others 
in identifying flavor multiplets 
and predicting the mass of the Omega
baryon suggests that the strange-quark mass is still light enough to allow for systematic calculations of 
flavor-symmetry breaking effects using chiral perturbation theory, i.e.\ performing perturbation theory in the three lightest 
quark masses. 
In particular, it appears rather unlikely that the {\it sign} of the $\Delta$-$\Delta$-$\pi$ coupling constant would turn 
out to be 
different from the sign of the $\Omega$-$\Xi^*$-$K$ coupling constant.\footnote{What is meant is always the relative sign 
between the axial-vector and the vector coupling constant for the various baryon combinations.} Under the assumption that 
the sign remains the same, the experiment can decide about the sign of $H_A$ by measuring the differential distribution of 
the decay (\ref{eq:Omega-weak}). This will be substantiated below by a crude leading-order calculation. For the decision about
the sign and the rough size this should be sufficient. 

\section{Research strategy}

The first step should be a measurement of the decay process \eqref{eq:Omega-weak}. 
But we have the QCD tools at hand to do much better than interpreting such a measurement by a crude leading-order calculation 
and assuming implicitly that the flavor-breaking effects are small when extrapolating to the Delta sector (where there is 
disagreement about the sign of the $\Delta$-$\Delta$-$\pi$ coupling constant). 
A better quantitative determination of $H_A$ and the impact of flavor breaking requires a dedicated theory initiative that 
complements the measurement of the process (\ref{eq:Omega-weak}). This provides a formidable task but is within reach. 
In general,  
processes like (\ref{eq:Omega-weak}) involve 14 form factors \cite{long-Omega-Xistar}. To deal with the quite involved tensor 
structures, it is suggestive to develop a 
projector formalism along the lines of \cite{Junker:2019vvy}. 

The general strategy for a model-independent determination 
of these form factors involves lattice QCD and loop calculations in chiral perturbation theory.
The latter contains several undetermined counter terms (low-energy constants beyond leading order) 
if one pushes the calculations to the loop level. 
In addition, the region of validity is limited to small momentum transfers and small quark masses. On the other hand, 
there are no problems in principle to address unstable states like the rather broad Delta baryons; 
see, e.g., \cite{Hilt:2017iup,Unal:2021byi}. Lattice QCD is not limited to particularly small quark masses or momentum 
transfers. Instead it is numerically expensive to use quark masses as light as the physical values for up- and down-quark masses.
In addition, the influence of excited states on the proper extraction of form factors must be carefully 
studied \cite{Bali:2018qus}. 
As already pointed out, it is highly non-trivial to address the form factors of broad resonances. This suggests to 
combine the two theory approaches. 

In contrast to experimental results, both theory approaches allow for a variation of the quark masses. In that way one can 
compare not only the dependence of the form factors on the momentum transfer but also on the quark masses. This allows for 
a more detailed investigation of the region where results from lattice QCD and from chiral perturbation theory agree. 
In this way one can use lattice-QCD results to pin down the values of low-energy constants. In a second step 
one can then use chiral perturbation theory to extrapolate to those decuplet states with smaller life times. 

Once the whole machinery has been developed, one can check to which degree the strength or even if the sign 
of the decuplet--decuplet--Goldstone-boson coupling constants depends on the strangeness sector. One can also check to which 
extent the Goldberger-Treiman relation is satisfied in the various strangeness sectors. 

But one should not blur the complications that lie on the way. The interplay between loop renormalization and power counting 
and its possible complications are well documented \cite{Scherer:2012xha}. 
In addition, calculations in chiral perturbation theory are often restricted
to the third chiral order \cite{Bauer:2012pv,Yao:2016vbz,Hilt:2017iup,Alvarado:2021ibw,Alvarado:2023loi}
in view of the plethora of low-energy constants --- 
in particular four-point interactions --- 
that appear at next-to-leading order (NLO) \cite{Holmberg:2018dtv}. Pushing the calculations to the fourth chiral order requires 
to fix these NLO low-energy constants, e.g.\ by fits to results from 
lattice QCD \cite{Ren:2013oaa,Lutz:2018cqo,Sauerwein:2021jxb,Alvarado:2023loi}. 
Of course, this calls also for a high quality on 
the lattice side, providing a variety of quark masses and momentum transfers for the form factors. 

Another topic of ongoing research is the question which power counting is most effective for calculations in chiral perturbation 
theory that include the Delta (two-flavor calculations) or the decuplet (three-flavor calculations) states. The decay region 
probed by \eqref{eq:Omega-weak} requires a power counting that is somewhat different from the one applicable to 
generic spacelike form factors and meson-baryon scattering reactions; 
see, e.g., the discussions in \cite{Jenkins:1991bt,Mommers:2022dgw}. In addition, there are different suggestions how to treat 
the mass difference between baryon octet and decuplet relative to the Goldstone boson masses ($M_K$ for kaons,
$M_\pi$ for pions). 
For three flavors, $m_{(10)}-m_{(8)} \sim M_K^2$ has been suggested in \cite{Jenkins:1991bt}, while $m_{(10)}-m_{(8)} \sim M_K$ 
has been used in \cite{Holmberg:2019ltw,Mommers:2022dgw}. For two flavors, the small-scale expansion $m_\Delta - m_N \sim M_\pi$ 
has been suggested in \cite{Hemmert:1997ye} while the ``delta expansion'' $(m_\Delta - m_N)^2 \sim M_\pi$ has been used 
in \cite{Pascalutsa:2002pi,Pascalutsa:2006up}. Comparisons between chiral perturbation theory and lattice QCD 
will shed more light on the most effective power counting scheme, but in view of these challenges, 
it should be clear that the necessary third pillar are high-quality data of the 
process \eqref{eq:Omega-weak} that serve to provide valuable constraints 
and therefore cross-checks of the theory results.

%
%

\section{Feasibility check}

To check the feasibility of experimental measurements, a leading-order calculation is presented for the 
reaction \eqref{eq:Omega-weak}. 
We stress again that eventually this should and can be improved by calculations in lattice QCD confronted with the corresponding 
(loop) calculations in chiral perturbation theory. The sole purpose of the present work is to motivate such activities in 
experiment and theory. 

In the following, $\Xi^*$ denotes the neutral spin-3/2 decuplet state with strangeness $-2$, $K$ denotes the charged kaon. 
The Lagrangian \eqref{eq:baryonlagr} leads to the following Feynman matrix element for the process \eqref{eq:Omega-weak}:
\begin{eqnarray}
  {\cal M} &=& \bar u_{\ell} \gamma^\alpha (\mathds{1}-\gamma_5) v_{\nu_\ell} \nonumber \\ && \times 
  \bar u_{\Xi^*}^\mu \left[
    c_V \gamma_\alpha - c_A \gamma^\beta \gamma_5 
    \left(
      g_{\alpha\beta} - \frac{q_\alpha q_\beta}{q^2- M_K^2}
    \right)
  \right] g_{\mu \nu} u_\Omega^\nu  \nonumber \\ &&
  \label{eq:matrel}  
\end{eqnarray}
with $q := p_\Omega - p_{\Xi^*}$ and 
\begin{eqnarray}
  c_V = \sqrt{\frac32} G_F V_{us} \,, \qquad c_A = H_A \sqrt{\frac16} G_F V_{us}
  \label{eq:cVcA-HA}
\end{eqnarray}
where $G_F$ denotes the Fermi
constant and $V_{us}$ the element of the Cabibbo-Kobayashi-Maskawa quark-mixing matrix \cite{pdg}. 

For massless leptons, the kaon pole in \eqref{eq:matrel} would not contribute. If one then replaced $H_A \to +3$, 
one would obtain an overall structure $\gamma_\alpha (\mathds{1}-\gamma_5)$, resembling the elementary structure of the weak 
interaction. Of course, spin-3/2 decuplet states are not elementary. A common way to construct interpolating quark currents 
for the spin-3/2 vector-spinors involves left- and right-handed quarks \cite{Ioffe:1981kw}. 
Therefore, a positive value for $H_A$ is not guaranteed.
But it makes it plausible that quark model and large-$N_c$ considerations lead to positive values for $H_A$. 

The complete angular distributions will be shown elsewhere \cite{long-Omega-Xistar}. Here we determine the decay width 
\begin{eqnarray}
  \Gamma_{\Omega \to \Xi^* \ell \bar\nu_\ell} 
  = \int\limits_{m_\ell^2}^{\Delta m^2} \! \! \! dq^2 \, \frac1{(2\pi)^3} 
  \frac{\vert \vec p_\ell\vert \, \vert \vec p_\Omega \vert}{16 m_\Omega^3} \int\limits_{-1}^{+1} d\cos\theta \;
  \overline{\vert {\cal M} \vert^2}
  \label{eq:widthint}  
\end{eqnarray}
and the forward-backward asymmetry 
\begin{eqnarray}
  \Gamma_{\rm fb}
  &:=& \int\limits_{m_\ell^2}^{\Delta m^2} \! \! \! dq^2 \, \frac1{(2\pi)^3} 
  \frac{\vert \vec p_\ell\vert \, \vert \vec p_\Omega \vert}{16 m_\Omega^3}  \nonumber \\ 
  && \times \left(
    \int\limits_{-1}^{0} d\cos\theta \; \overline{\vert {\cal M} \vert^2} - 
    \int\limits_{0}^{+1} d\cos\theta \; \overline{\vert {\cal M} \vert^2} 
  \right)
  \label{eq:widthfbint}  
\end{eqnarray}
where $\Delta m := m_\Omega-m_{\Xi^*}$. 
Results are presented 
for the cases $H_A = \pm 2,0$ for electrons and for muons. 
In the previous formulae, $q^2$ denotes the square of the invariant mass
of the dilepton system. The angle $\theta$ and the three-momenta are determined in the center-of-mass frame of the dilepton 
(not in the rest frame of Omega). The angle $\theta$ is measured between the flight direction of the charged 
lepton $\ell$ and the baryon direction.\footnote{In this dilepton rest frame, both baryons fly in the very same direction.} 

The results are provided in table \ref{tab:res}. 
\begin{table}[h]
  \centering
  \begin{tabular}{|l||c|c|}
    \hline & $\Gamma_{\Omega \to \Xi^* \ell \bar\nu_\ell}/\Gamma_{\Omega,\rm tot}$ & $\Gamma_{\rm fb}/\Gamma_{\Omega \to \Xi^* \ell \nu}$ \\ \hline
    $\ell = e$, $H_A = +2$ & $1.2 \cdot 10^{-4}$ & $+0.011$ \\ \hline
    $\ell = e$, $H_A = 0$ & $6.7 \cdot 10^{-5}$ & $-0.00043$  \\ \hline
    $\ell = e$, $H_A = -2$ & $1.2 \cdot 10^{-4}$ & $-0.012$ \\ \hline
    $\ell = \mu$, $H_A = +2$ & $4.3 \cdot 10^{-6}$ & $-0.23$ \\ \hline
    $\ell = \mu$, $H_A = 0$ & $2.5 \cdot 10^{-6}$ & $-0.33$ \\ \hline
    $\ell = \mu$, $H_A = -2$ & $4.3 \cdot 10^{-6}$ & $-0.25$ \\ \hline
  \end{tabular}
  \caption{Branching ratios and forward-backward (fb) asymmetry for various cases.}
  \label{tab:res}
\end{table}
To interpret these results, it is important to understand that for massless leptons the forward-backward asymmetry is 
proportional to the product $c_V \cdot c_A$. For massive leptons, there are extra terms that scale with $c_V^2$ and $c_A^2$; see 
also \cite{Holmberg:2019ltw,Mommers:2022dgw} for a related discussion. 
The decay involving an electron is sufficiently close to the massless case, while the muon is not. 
The results suggest that the decay with an electron is well suited to determine the sign of $H_A$. But one needs to achieve 
a 1\% accuracy to resolve the forward-backward difference, i.e.\ one needs sufficient statistics for a high-quality 
angular distribution. 

For the decay with a muon, 
the differences caused by a sign change of $H_A$ are small. It remains to be seen if 
a fully differential distribution reveals more differences \cite{long-Omega-Xistar}. But in addition, the muon case has 
a significant phase-space suppression. Thus the electron case appears to be more promising. 

The estimates of the branching ratio indicate that the decay \eqref{eq:Omega-weak} is rare but should be within reach of present 
experiments like BESIII \cite{BESIII:2020lkm} or LHCb \cite{Belyaev:2021cyr}. 
The branching ratio does not depend on the product $c_V \cdot c_A$. It contains only $c_V^2$ and $c_A^2$ terms. Thus it cannot be
used to determine the sign of $H_A$. But if the branching ratio can be determined with a precision of, say, 10\%, then it will 
be possible to estimate the size of the low-energy constant $H_A$. 
If the flavor-breaking effect is comparable to the case of the other 
decuplet related low-energy constant $h_A$ (around $\pm 20$\%), then the measurement of the process \eqref{eq:Omega-weak} 
will provide a first direct 
determination of size and possibly sign of the last unknown low-energy constant of the leading-order Lagrangian of chiral 
perturbation theory. 

\bibliography{lit}{}
\bibliographystyle{apsrev4-1}

\end{document}